\documentclass[english,aps,prd,preprint,nofootinbib]{revtex4-2}
\usepackage{mathrsfs}
\usepackage[T1]{fontenc}
\usepackage[utf8]{inputenc}
\usepackage{graphicx}
\usepackage{esint}
\usepackage{babel}
\usepackage{color}

\makeatletter
\@ifundefined{textcolor}{}
{%
 \definecolor{BLACK}{gray}{0}
 \definecolor{WHITE}{gray}{1}
 \definecolor{RED}{rgb}{1,0,0}
 \definecolor{GREEN}{rgb}{0,1,0}
 \definecolor{BLUE}{rgb}{0,0,1}
 \definecolor{CYAN}{cmyk}{1,0,0,0}
 \definecolor{MAGENTA}{cmyk}{0,1,0,0}
 \definecolor{YELLOW}{cmyk}{0,0,1,0}
}

\makeatother

\begin{document}

\preprint{This line only printed with preprint option}

\title{Free energy, stability, and particle source in dynamical holography}

\author{Yu Tian}
\email{ytian@ucas.ac.cn}
\affiliation{School of Physical Sciences, University of Chinese Academy of Sciences, Beijing 100049, China}
\affiliation{Institute of Theoretical Physics, Chinese Academy of Sciences, Beijing 100190, China}

\author{Xiao-Ning Wu}
\email{wuxn@amss.ac.cn}
\affiliation{Institute of Mathematics, Academy of Mathematics and System Science, Chinese Academy of Sciences, Beijing 100190, China}
\affiliation{Hua Loo-Keng Key Laboratory of Mathematics, Chinese Academy of Sciences, Beijing 100190, China}

\author{Hongbao Zhang}
\email{hongbaozhang@bnu.edu.cn}
\affiliation{Department of Physics, Beijing Normal University, Beijing 100875, China}

\begin{abstract}
By using the conserved currents in the bulk space-time, we study dynamical holographic systems and the relation between thermodynamical and dynamical stability of such systems. In particular, in the probe limit a generalized free energy is defined with the property of monotonic decreasing in dynamic processes. It is then shown that the (absolute) thermodynamical stability implies the dynamical stability, while the linear dynamical stability implies the thermodynamical (meta-)stability. The holographic superfluid is taken as an example to illustrate our general formalism, where the dynamic evolution of the system in contact with a particle source is clarified by theoretical investigation and numerical verification. The case going beyond the probe limit is also discussed.
\end{abstract}

\maketitle

\section{{Introduction}}

Real systems in our world are complicated systems in dynamical evolution. But we always need approximation, simplification and modeling when doing physics. If some systems are evolving in time and varying in space slowly enough, we can model them as (locally) static, homogeneous systems, i.e. systems in (local) equilibrium, where thermodynamics (hydrodynamics) applies. As a step further, if the systems are just slightly perturbed away from local equilibrium, we can phenomenologically describe them by linear response theory, where the systems are characterized by various kinds of response functions in addition to thermodynamic quantities. If the systems are really far from local equilibrium, then in principle the microscopic details of the systems matter, i.e. we have to use the full quantum field/many-body theory to describe the dynamical quantum many-body systems, which is extremely difficult (if ever possible).\footnote{Note that a system could be far away from global equilibrium while still preserving local equilibrium, which is just the occasion for the application of hydrodynamics. Further, even for some systems far from local equilibrium certain modeling/approximations can be used to obtain meaningful dynamical descriptions, which usually have great limitations in turn.} However, for quantum many body systems in the holographic limit, i.e. with a gravity dual, we can use AdS/CFT (holography) to study the full dynamics of them (see, e.g. \cite{Sonner} for a recent review).

Although dynamical holography is very interesting and strongly motivated and already has significant achievements, it is not systematically justified, compared to the standard AdS/CFT in the equilibrium (Euclidean) case\cite{AdS-CFT,Witten,GKP} and its Minkowski extension in the linear response regime\cite{Minkowski}. But as long as it is self-consistent and compatible with fundamental physical theories (like hydrodynamics) and the standard AdS/CFT in the (near-)equilibrium limit, and it has prediction power, it is important and valuable at least in the phenomenological sense and we should try to make it as systematic as possible. Generally, dynamical processes in holography are studied using numerics\cite{Sonner}, but it would be very helpful that we could have some quantities to control general dynamical processes analytically, which is the main goal of this paper.

An important problem closely related to non-equilibrium physics is stability, in either the linear level or the nonlinear level. The correlated stability, namely, the equivalence between thermodynamical and dynamical stability in gravitational physics, is a long-standing problem and has recently received renewed attention due to AdS/CFT correspondence. Particularly, in AdS/CFT the bulk spacetime with a black hole is dual to the boundary system at a finite temperature. So it is tempting to conjecture that the correlated stability should hold for such holographic gravitational systems\cite{GM1,GM2}. But whether this conjecture turns out to be true or not, it is meaningful to consider it in the context of non-equilibrium physics in holographic systems.

In this paper we use the conserved current method developed in \cite{Wald1} to study dynamical holographic systems and the relation between thermodynamical and dynamical stability of such systems. The case with fixed spacetime backgrounds (in the probe limit, i.e. ignoring the back reaction of bulk matter fields onto the background spacetime) is discussed first in Sec.~\ref{probe}, where a generalized free energy is defined with the property of monotonic decreasing in dynamic processes. That important quantity is then used to argue the relation between thermodynamical and dynamical stability. We take the simplest holographic superfluid model as an important example in Sec.~\ref{superfluid}, where the relevant quantities are explicitly calculated, and different types of dynamical evolutions are discussed in the view of the conserved currents.

The case with full back-reaction is much more complicated, as discussed in Sec.~\ref{back-reaction}, where we generalize part of the previous investigation with the help of conserved currents associated to the diffeomorphism invariance. In both cases, our analyses help to clarify some aspects of the far-from-equilibrium holographic physics. We end our paper with some discussion and outlook in Sec.~\ref{discuss}.

\section{Conserved currents and free energies in the probe limit}\label{probe}

Let us start with the Lagrangian density of a collection of matter fields $\phi$ on top of a fixed background (bulk black hole) metric $g_{ab}$:
\begin{equation}\label{Lagrangian}
	\mathbf{L}=\mathscr{L}(g_{ab}, \phi, \nabla_a\phi)\epsilon,
\end{equation}
where $\nabla$ and $\epsilon$ are the derivative operator and volume element compatible with $g$, respectively. The variation of Lagrangian density gives
\begin{equation}\label{var_L}
	\delta\mathbf{L}=\frac{1}{2}T^{ab}\epsilon\delta g_{ab}+E\epsilon\delta\phi+d\mathbf{\Theta}=\frac{1}{2}T^{ab}\epsilon\delta g_{ab}+d\mathbf{\Theta},
\end{equation}
where $T^{ab}=2\frac{\delta\mathscr{L}}{\delta g_{ab}}+\mathscr{L}g^{ab}$ is the energy momentum tensor for the matter fields, $\mathbf{\Theta}=\frac{\partial\mathscr{L}}{\partial\nabla_a\phi}\cdot\epsilon\delta\phi$, and the equation of motion $E=\frac{\partial\mathscr{L}}{\partial\phi}-\nabla_a\frac{\partial\mathscr{L}}{\partial\nabla_a\phi}=0$ has been used in the second step. Now let us focus on such a variation induced by the infinitesimal diffeomorphism generated by a vector field $\xi$, namely $\delta g_{ab}=\mathcal{L}_\xi g_{ab}=\nabla_a\xi_b+\nabla_b\xi_a$ and $\delta\phi=\mathcal{L}_\xi\phi$. Then the diffeomorphism covariance of our Lagrangian density gives rise to the following identity
\begin{equation}\label{identity}
	d(\mathbf{J}_\xi-\mathbf{K}_\xi)=\xi_b\nabla_a T^{ab}\epsilon,
\end{equation}
where
\begin{equation}
	\mathbf{J}_\xi=\mathbf{\Theta}_\xi-\xi\cdot\mathbf{L}=j_\xi\cdot\epsilon,\quad \mathbf{K}_\xi=k_\xi\cdot\epsilon
\end{equation}
with $j_\xi^a=\frac{\partial\mathscr{L}}{\partial\nabla_a\phi}\mathcal{L}_\xi\phi-\xi^a\mathscr{L}$ and $k_\xi^a=-T^{ab}\xi_b$. {Since $\xi$ is arbitrary,} this identity implies the conservation of energy momentum tensor $\nabla_a T^{ab}=0$, which further leads to the equivalence of Noether current $j_\xi$ and stress energy current $k_\xi$ in the sense of $\mathbf{K}_\xi=\mathbf{J}_\xi+d\mathbf{B}$.

{It is noteworthy that on many occasions we need to add boundary terms to the Lagrangian (as will be discussed in the following sections), where we have
\begin{equation}\label{boundary}
\tilde{\mathbf{L}}=\mathscr{L}(g_{ab}, \phi, \nabla_a\phi)\epsilon+d\mathbf{b}(g_{ab},\phi,\nabla_{a}\phi)
\end{equation}
instead of (\ref{Lagrangian}). Then there will be
\begin{equation}
\delta\tilde{\mathbf{L}}=\frac{1}{2}T^{ab}\epsilon\delta g_{ab}+d\mathbf{\Theta}+d\delta\mathbf{b}
\end{equation}
again by virtue of the equations of motion, which remain the same form as in (\ref{var_L}) in spite of $\mathbf{b}$. The diffeomorphism invariance with respect to an arbitrary $\xi$ again tells us
\begin{equation}
d(\tilde{\mathbf{J}}_{\xi}-\mathbf{K}_{\xi})=\xi_{b}\nabla_{a}T^{ab}\epsilon=0
\end{equation}
with the modified Noether current
\begin{equation}
\tilde{\mathbf{J}}_{\xi}=\mathbf{\Theta}_{\xi}+\mathcal{L}_{\xi}\mathbf{b}-\xi\cdot\tilde{\mathbf{L}}.
\end{equation}
One sees that the difference
\begin{equation}
\tilde{\mathbf{J}}_{\xi}-\mathbf{J}_{\xi}=\mathcal{L}_{\xi}\mathbf{b}-\xi\cdot d\mathbf{b}=d(\xi\cdot\mathbf{b}),
\end{equation}
which explicitly shows the equivalence of $\tilde{\mathbf{J}}_{\xi}$ and $\mathbf{J}_{\xi}$. Thus, we will not take into account the effect of possible boundary terms until Sec.~\ref{back-reaction}, where it is necessary. Note that the stress energy current $\mathbf{K}_{\xi}$ will never get influenced by the addition of $\mathbf{b}$.} For a Killing vector field $\xi$, both currents are closed, namely $d\mathbf{J}_\xi=d\mathbf{K}_\xi=0$. This is essentially the Noether theorem for the spacetime symmetries. Under the probe limit, we shall focus on the conserved currents with $\xi=\partial_t$ the future directed Killing vector field, by which the black hole horizon is generated.

For an equilibrium state ($\mathcal{L}_t\phi=0$), the flux across either the black hole horizon or AdS boundary for $\mathbf{J}_t$ apparently vanishes. So does the flux for $\mathbf{K}_t$. Therefore the flux across any surface bounded by the black hole horizon and AdS boundary is equal to the flux across the surface $\Sigma$ of equal time, which is obviously the free energy $-\int_\Sigma\sqrt{-g}\mathscr{L}$ for $\mathbf{J}_t$. Actually, as will be seen more clearly from the example in the next section, for systems involving gauge fields $\mathbf{J}_t$ will be gauge dependent while $\mathbf{K}_t$ is manifestly gauge invariant. There we can learn that for those systems the flux for $\mathbf{K}_t$ is the genuine free energy, while that for $\mathbf{J}_t$ is actually the grand potential associated to the conserved global charge of the boundary system (dual to the conserved local charge corresponding to the gauge symmetry in the bulk). Thus, we shall discuss solely the stress energy current $k_\xi$ in the rest of this section and come back to the Noether current $j_\xi$ later.

For a state out of equilibrium but varying slowly enough in space and time,\footnote{Similar requirements are usually called adiabatic in the literature, but literally ``adiabatic" only emphasizes slow temporal variation.} the free energy and other thermodynamic quantities can be well defined locally, as well as the thermodynamic relations can hold locally. In this case, the system is in \emph{local equilibrium}, which can be described by non-equilibrium thermodynamics or hydrodynamics\cite{Reichl}. Away from local equilibrium, i.e. the system is in a state varying rapidly in either space or time, the thermodynamic or hydrodynamic description breaks down, and in general there is no well-defined free energy or other thermodynamic quantities.

However, at least for the holographic systems in the probe limit that we are considering, a natural generalization of free energy can be well defined using the conserved current $k_t$, even far from (local) equilibrium. Actually, we just go on to define the (generalized) free energy at time $t$ as the flux of $k_t$ across the constant time $t$ slice $\Sigma_t$:
\begin{equation}
F(t):=\int_{\Sigma_t}\mathbf{K}_t.
\end{equation}
It is then easy to show that $F(t)$ defined above has the following properties:
\begin{enumerate}
	\item It becomes the standard free energy in the local equilibrium limit;
	\item It decreases monotonically in a general dynamical process (no need to be in local equilibrium) without external work (or called driving\cite{driving}) done to the boundary system;
	\item Its decrease exactly matches the integral of the energy flux across the horizon (see \cite{vortex,TWZ1,TWZ2}, for example).
\end{enumerate}
The first property is just trivial by definition. We shall prove the last two properties in the rest of this section.

By the Stokes theorem, the total flux of this current across the boundary of any given spacetime region always vanishes. For our purpose, we would like to focus on the region sandwiched by two slices, denoted by $\Sigma_i$ and $\Sigma_f$ respectively. This spacetime region also has the black hole horizon $H$ as the inner boundary and the conformal infinity as the outer AdS boundary. The flux across the AdS boundary just corresponds to the external work done to the boundary system (see \cite{TWZ1,TWZ2}, for example), which vanishes according to our premise of this discussion. With this in mind, we have the following equality
\begin{equation}
	\int_{\Sigma_i}\mathbf{K}_t=\int_{\Sigma_f}\mathbf{K}_t+\int_H\mathbf{K}_t.
\end{equation}
Now come two important observations. First, if the system is in equilibrium on the slice $\Sigma$, then the flux across this slice is exactly the usual free energy, as mentioned above.
Second, the flux across the black hole horizon is believed to be always positive due to the null energy condition. Then it follows that the dynamical evolution leads to the decrease of its free energy, where the decrease is the energy dissipation.

{Therefore, if an equilibrium state is thermodynamically stable in the absolute sense, i.e. it has a free energy that is a global minimum in the phase space, then it should be dynamically stable as well, because otherwise the system would eventually be driven to an equilibrium state with an even lower free energy. Conversely, if the equilibrium state in consideration is not thermodynamically (meta-)stable, i.e. with a free energy that is not a local minimum, then there exist tiny perturbations on top of such a state that can trigger the (irreversible) dynamical evolution that the free energy of the system rolls down, because in thermodynamics the instability shows up under certain inhomogeneous thermodynamic perturbations as discussed in standard textbooks, which are just the long-wavelength limit of the corresponding perturbations in the holographic system. In other words, the linear dynamical stability implies the thermodynamical (meta-)stability.}

Here is a remark. {In the above derivation, we have ignored the holographic renormalization procedure generically required for the holographic set-up, which is related to the possible boundary terms considered in (\ref{boundary}) and will be discussed in Sec.~\ref{back-reaction}.}

\section{The holographic superfluid as an example}\label{superfluid}

Among others, holographic superfluid\cite{HHH1} is supposed to be the prototype of holographic condensed matter systems in applied AdS/CFT. So here we would like to take it as a concrete example to demonstrate the validity of the main claims in our proof. The fixed background geometry is the Schwarzschild planar black hole, which can be written in the Eddington-Finkelstein coordinates as
\begin{equation}
	ds^2=\frac{L^2}{z^2}(-fdt^2-2dtdz+dx^2+dy^2)
\end{equation}
with $f=1-({z}/{z_0})^3$, $L$ the AdS curvature radius and $z_0$ the black hole horizon radius. The bulk Lagrangian for holographic superfluid is simply the Maxwell field coupled to a massive complex scalar field, given by
\begin{equation}\label{HHH}
	\mathscr{L}=-\frac{1}{4}F_{ab}F^{ab}-|D\psi|^2-m^2|\psi|^2,
\end{equation}
where $D=\nabla-iA$ with $A$ the gauge potential, and $m^2L^2=-2$ for simplicity. The asymptotic behavior for the matter fields goes as
\begin{equation}\label{asymptotic}
	A_\nu=a_\nu+b_\nu z+o(z),\quad\psi=\psi_0 z+\psi_1z^2+o(z^2),
\end{equation}
where $\nu$ denotes the AdS boundary coordinates with $a_t$ and $-b_t$ the chemical potential $\mu$ and charge density $\rho$ of the boundary system by holography in the gauge $A_t|_{z_0}=0$ and $A_z=0$. {As the simplest case for illustration, in the standard quantization of the holographic superfluid model (see, e.g. \cite{LanTZ} for a discussion of dynamic evolution in different quantizations), $\psi_0$ is taken as the source and is set to zero when there is no driving.} The detailed scheme for the dynamic evolution, in particular in the inhomogeneous case, can be seen in, e.g \cite{DNTZ}. 

The energy momentum tensor is given by
\begin{equation}
	T^{ab}=\frac{2}{\sqrt{-g}}\frac{{\delta I}}{\delta g_{ab}}=F^a{}_cF^{bc}+D^a\psi({D^b\psi})^*+D^b\psi({D^a\psi})^*+\mathscr{L}g^{ab}.
\end{equation}
The associated conserved current can be obtained as
\begin{equation}\label{current_diff}
	k_\xi^a=j_\xi^a+\nabla_b(F^{ab}A_c\xi^c),
\end{equation}
which can also be expressed as
\begin{equation}
	\mathbf{K}_\xi=\mathbf{J}_\xi+d*(\mathbf{F}A_c\xi^c)
\end{equation}
with $*$ the Hodge dual. The flux of the last term in (\ref{current_diff}) across the boundary can be calculated as
\begin{eqnarray}
n_{b}\nabla_{a}(F^{ab}A_{\xi})&=&\nabla_{a}(n_{b}F^{ab}A_{\xi})-F^{ab}A_{\xi}\nabla_{a}n_{b}\nonumber\\
&=&\nabla_{a}(n_{b}F^{ab}A_{\xi})+n_{b}F^{ab}A_{\xi}n^{c}\nabla_{c}n_{a}\nonumber\\
&=&(\delta_{a}^{c}-n_{a}n^{c})\nabla_{c}(n_{b}F^{ab}A_{\xi})\nonumber\\
&=&\bar{\nabla}_{a}(n_{b}F^{ab}A_{\xi})=\bar{\nabla}_{a}(j^{a}A_{\xi})\label{flux_diff}
\end{eqnarray}
with $A_\xi:=A_c\xi^c$, $n^a$ the unit normal to the boundary, $\bar{\nabla}_{a}$ the induced derivative on the boundary and $j^a=n_b F^{ab}$ the particle current on the boundary, while the flux of the stress energy current is
\begin{equation}\label{flux}
	n_a k^a_\xi=j^a F_{ba}\xi^b+\Pi\xi^a D_a\psi+\Pi^*(\xi^a D_a\psi)^*.
\end{equation}
Here $\Pi=n_a D^a\psi^*$ is the conjugate momentum to $\psi$ (with respect to the holographic direction instead of time).

Note that the flux of the Noether current is gauge dependent, but in the gauge $A_t|_{z_0}=0$ the fluxes of the two conserved currents at the horizon coincide, so the flux of the Noether current there is also positive definite in the dynamical case and then has the meaning of dissipation. Together with $A_z=0$ in the bulk, the gauge $A_t|_{z_0}=0$ also fixes $a_t$ in (\ref{asymptotic}), which can be defined as the generalized chemical potential $\mu$ in the dynamical case. As we will see even in the back-reacted case, these gauge conditions are natural and preferential for general holographic systems involving bulk gauge fields, which will be insisted on in the rest of this paper.

{Regarding the holographic renormalization, the required counter term $-\frac{1}{L}\int_B\sqrt{-h}|\psi|^2$ for the case of standard quantization considered here does not contribute the free energy due to the aforementioned source free boundary condition.} In particular, by virtue of (\ref{flux_diff}), the flux of the stress energy current corresponds to the free energy ${\int_\Sigma(-\sqrt{-g}\mathscr{L}+\rho\mu)}$ in the canonical ensemble and the flux of the Noether current corresponds to the free energy (grand potential) $-\int_\Sigma\sqrt{-g}\mathscr{L}$ in the grand canonical ensemble. For a relaxation process, i.e. dynamical evolution without driving in our holographic superfluid, besides $\psi_0=0$, we are also required to set $F_{\nu\sigma}|_0=0$, since all sources should be turned off.

\subsection{{Dynamical evolution under the condition of fixed particle number or fixed chemical potential}}\label{canonical_dynamics}

In dynamical time evolution, the total particle number of the holographic superfluid can be fixed or the system can be put into contact with a particle source with a fixed chemical potential. Naively, one may think that these two cases can be distinguished by different boundary conditions at the AdS conformal boundary. Similar to the equilibrium case, we may call the corresponding holographic boundary conditions the ``canonical'' and ``grand canonical'' boundary conditions, respectively.

The ``canonical'' boundary condition is simple. For a system with fixed particle number and no external work (in relaxation or without driving), the flux of the energy current $\mathbf{K}_t$ across the conformal boundary (corresponding to the external work\cite{TWZ2})\footnote{See also the discussion in Sec.~\ref{back-reaction}.} vanishes, as well as the flux
\begin{equation}\label{divergence}
	-i\Pi{\psi}+i\Pi^*{\psi}^*=\bar{\nabla}_\mu j^\mu
\end{equation}
of the bulk electric current $J_{E}$ (corresponding to the variation of the particle number)\cite{TWZ2}. Obviously, either Dirichlet or Neumann boundary condition for $\psi$ fixes the total particle number.

But the ``grand canonical'' boundary condition is problematic. For a system with fixed chemical potential $\mu$ and no external work, the flux of the Noether current $\mathbf{J}_t$ across the conformal boundary (supposed to correspond to the external work under fixed chemical potential) vanishes, as well as the boundary value of $A_t$ should be fixed. Because of (\ref{current_diff}) and (\ref{flux_diff}), the local fluxes of those two currents have the relation
\[ n_a k_t^a=n_a j_t^a+\bar{\nabla}_\mu (\mu j^\mu)=n_a j_t^a+\mu\bar{\nabla}_\mu j^\mu, \]
so the vanishing of the Noether flux $n_a j_t^a$ means
\begin{eqnarray}
	0&=&n_a k_t^a-\mu\bar{\nabla}_\mu j^\mu\nonumber\\
	&=&j^a F_{ta}+\Pi\bar{\nabla}_t\psi+\Pi^*\bar{\nabla}_t\psi^*,\label{grand}
\end{eqnarray}
where we have used (\ref{flux}) and (\ref{divergence}). The first term on the right hand side of (\ref{grand}) is just the work done by an external potential (with the corresponding force $F_{ta}$). If we do not want to turn on the external potential, the boundary condition (besides $\mu$ fixed) becomes
\begin{equation}
	\Pi\bar{\nabla}_t\psi+\Pi^*\bar{\nabla}_t\psi^*=0,
\end{equation}
which is also satisfied by either Dirichlet or Neumann boundary condition for $\psi$. However, if we impose the boundary conditions $A_t|_{z_0}=0$ (gauge fixing) and $A_t|_{0}=\mu$ (chemical potential fixing) for $A_t$ simultaneously, there is no room to require the constraint equation (\ref{divergence}) as an additional boundary condition.

Therefore, we see that the condition of fixed chemical potential cannot be imposed in the above way. Instead, we should really put the system into contact in spatial directions with an environment (particle source). In other words, the difference between the canonical case and the grand canonical case only shows up for systems with spatial boundaries. Note that here and in the following the radial (holographic) direction $z$ is not viewed as a ``spatial'' direction.

So now the problem is shifted to how to impose boundary conditions for holographic systems with spatial boundaries. For the ``canonical'' case, the appropriate boundary conditions at, say, the $x$ boundary are
\begin{equation}\label{x_boundary}
	F^{xz}=0,\qquad F^{xy}=0,\qquad D_x\psi=0
\end{equation}
for $A_x$, $A_y$ and $\psi$, with no extra boundary condition imposed there for $A_t$. These boundary conditions guarantee $j^x=n_b F^{xb}|_0=0$ at the $x$ edge of the conformal boundary, so there is no flux of the particle current across the $x$ edge.\footnote{Actually, these boundary conditions also guarantee that there is no flux of the bulk electric current $J_E$ across the $x$ boundary.} {Most importantly, the flux $k_t^x$ of the energy current $\mathbf{K}_t$ across the $x$ boundary vanishes, because
\begin{equation}\label{x_flux}
	k_t^x=-T_t^x=-F_{ta} F^{xa}-D_t\psi (D^x\psi)^*-D^x\psi (D_t\psi)^*=0,
\end{equation}
which guarantees the monotonic decrease of the free energy in relaxation.} More explicitly for numerical schemes of time evolution, the above boundary conditions are
\begin{equation}
	\partial_t A_x=\partial_x A_t+f\partial_z A_x,\qquad \partial_x A_y=\partial_y A_x,\qquad\partial_x\psi=i A_x\psi.
\end{equation}
The discussion about the $y$ boundary is similar.

{On some occasions, the boundary condition $F^{xy}=0$ in (\ref{x_boundary}) may not be possible.\footnote{For example, the rotating holographic superfluid discussed in \cite{LTZ}.} Then we can use the boundary condition
\begin{equation}
F_{ty}=0
\end{equation}
instead, which means explicitly
\begin{equation}
\partial_t A_y=\partial_y A_t.
\end{equation}
Obviously, this boundary condition together with the other two conditions in (\ref{x_boundary}) also guarantees (\ref{x_flux}).}

For the ``grand canonical'' case, instead, the following boundary conditions should be used at the $x$ boundary:
\begin{equation}
	\partial_x A_x=0,\qquad A_y=0,\qquad\psi(z)=\psi_H(z)
\end{equation}
with $\psi_H(z)$ the homogeneous configuration of $\psi$ at the given chemical potential $\mu$. From the boundary evolution equation\cite{DLTZ}
\begin{equation}
	\partial_{t}(\partial_{z}A_{t}+\partial_{j}A_{j})-\partial_{j}\partial_{z}A_{j}=0
\end{equation}
of the particle number density $\rho$, the above boundary conditions conserve $\rho$ at the $x$ boundary (as its homogeneous value at the given $\mu$). Then from the constraint equation
\begin{equation}
	\partial_{z}(\partial_{z}A_{t}-\partial_{j}A_{j})=i(\psi^{*}\partial_{z}\psi-\psi\partial_{z}\psi^{*}),
\end{equation}
the above boundary conditions (together with $A_t|_{z_0}=0$ and $A_t^\prime|_0=-\rho$) just make $A_t(z)$ there to be the homogeneous configuration at the given $\mu$.

\begin{figure}
	\begin{centering}
		\includegraphics[scale=0.6]{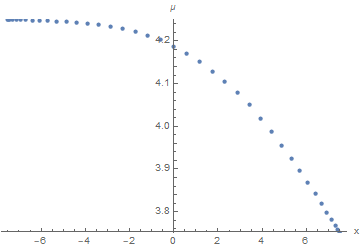}
	\end{centering}
	\caption{Profile of the (generalized) chemical potential $\mu$ in the steady state with a persistent particle current going from the left end ($\mu_2=4.25$) to the right end ($\mu_1=3.75$), after evolution for a long time.\label{fig:profile}}
\end{figure}

To illustrate and verify the above scheme, here we show the result of a numerical simulation, mimicking the dynamics of the superfluid system (\ref{HHH}) in relaxation between two particle sources with different chemical potentials. Specifically, we set $\mu_1=3.75$ and $\mu_2=4.25$ (dimensionless unit with $L=1$ and $z_0=1$) so that the region near the second particle source is in a superfluid state while the rest is (roughly speaking) in a normal state, since the critical chemical potential for the superfluid transition is $\mu_c\approx 4.07$ in the standard quantization that we chose. These two sources are separated by a distance of 15 in the $x$-direction. After evolution for a long time, the system approaches a steady state, the chemical potential profile of which is shown in Fig.~\ref{fig:profile}, with a persistent particle current $j^x\approx 0.086$ going from the second particle source (with a higher chemical potential) to the first.

\section{Beyond the probe limit: dynamical holographic spacetime}\label{back-reaction}

Even in the case with back-reaction, we can have the conserved currents associated to the diffeomorphism invariance, similar to that in Sec.~\ref{probe}. The main difference, though, is that in the dynamical case the bulk spacetime generally does not have a timelike Killing vector field. Again aiming at the holographic superfluid model\cite{HHH2}, we consider the Einstein gravity minimally coupled to a charged scalar field $\Phi$ and a Maxwell field $A_{\mu}$,
while keeping in mind that the discussion here can be well applied to general minimally coupled gravity-matter theories. The action is
\begin{eqnarray}
I & = & \int\mathscr{L}\sqrt{-g}d^{d+1}x,\label{eq:action}\\
\mathscr{L} & = & \frac{1}{16\pi G}(R-2\Lambda)-\frac{1}{2}\nabla_{\mu}A_{\nu}F^{\mu\nu}-(\nabla_{\nu}-iA_{\nu})\Phi(\nabla^{\nu}+iA^{\nu})\Phi^{*}.\nonumber
\end{eqnarray}
As discussed in \cite{Wald1}, the above theory has the conserved (Noether)
current\footnote{Note that in this papar we take $g_{\mu\nu}$ as the fundamental variables, so $\mathcal{L}_{\xi}g^{\mu\nu}$ here means the operation on $g_{\mu\nu}$ followed by levitation of the tensor indices with the metric.}
\begin{equation}
J_\xi^{\mu}=\frac{1}{16\pi G}(\nabla_{\nu}\mathcal{L}_{\xi}g^{\mu\nu}-\nabla^{\mu}\mathcal{L}_{\xi}g_{\nu}^{\nu})+\frac{\partial\mathscr{L}}{\partial\nabla_{\mu}\phi^{A}}\mathcal{L}_{\xi}\phi^{A}-\mathscr{L}\xi^{\mu}\label{eq:current}
\end{equation}
associated to the diffeomorphism invariance of $I$ induced by an
arbitrary vector field $\xi$, where $\phi^{A}$ runs over components
of all matter fields. {Actually, one may also replace the matter contribution in the last two terms of the above current with $-T^{ab}\xi_b$ to obtain another conserved current
\begin{equation}\label{K_xi}
K_\xi^\mu=\frac{1}{16\pi G}(\nabla_{\nu}\mathcal{L}_{\xi}g^{\mu\nu}-\nabla^{\mu}\mathcal{L}_{\xi}g_{\nu}^{\nu})-T^{\mu\nu}\xi_\nu-\frac{1}{16\pi G}(R-2\Lambda)\xi^{\mu},
\end{equation}
corresponding to the stress energy current in Sec.~\ref{probe}.}

\begin{figure}
\begin{centering}
\includegraphics[scale=0.6]{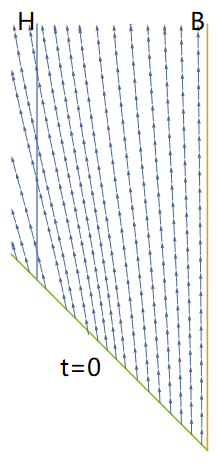}
\end{centering}
\caption{A cartoon of our configuration. The vectors represent the conserved currents.\label{fig:configure}}
\end{figure}

Our configuration is as Figure \ref{fig:configure}. There is a
horizon\footnote{It has long been discussed in the context of holography whether the event horizon or the apparent horizon is related to the entropy of the boundary system in the fully dynamic case, or even whether such an entropy can be well defined far from equilibrium. This problem remains open here.} $H$ in the (dynamic) space-time, which tends to the asymptotic
Killing horizon of the final asymptotic stationary black hole. There
is also a time-like boundary surface $B$, either at a finite distance
or tending to the conformal infinity, which together with the event
horizon composes the boundary of the bulk space-time. Here we illustrate the configuration with the characteristic dynamic evolution (see, e.g. \cite{Chesler,NCTWZ}), the most commonly used scheme in holography, so the initial state at $t=0$ is on a null surface.

Since the current (\ref{eq:current}) is conserved, we have
\begin{equation}\label{eq:Stokes}
0=\int_{{\rm bulk}}\nabla_{\mu}J_\xi^{\mu}=\int_{\Sigma_{i}}m_{\mu}J_\xi^{\mu}-\int_{\Sigma_{f}}m_{\mu}J_\xi^{\mu}+\int_{B}n_{\mu}J_\xi^{\mu}-\int_{H}\lambda_{\mu}J_\xi^{\mu},
\end{equation}
where $\lambda_{\mu}$ is {future directed and tangent to the null
geodesic generators of $H$ if $H$ is null (otherwise it is the unit normal vector), $m_{\mu}$ is the future directed unit normal vector to a time slice $\Sigma$, and $n_{\mu}$ the outward unit normal vector
to the time-like boundary surface $B$. In the discussion of decay of an unstable equilibrium state, we should assume that in the
infinite past, as well as in the infinite future, $\xi$ tends to the corresponding
asymptotic Killing vector field of the asymptotic stationary black
hole. In this case, only the last two terms in the conserved current (\ref{eq:current}) do not vanish, so it just becomes the conserved current $J_\xi$ in Sec.~\ref{probe}\footnote{But note that now the Lagrangian density $\mathscr{L}$ includes the contribution from gravity.} and its flux across a time slice is the generalized grand potential.} Correspondingly, the stress energy current (\ref{K_xi}) leads to the generalized free energy, the difference of which with the generalized grand potential is the same as in the previous discussion. In the following we just focus on the Noether current $J_\xi$, with the discussion about $K_\xi$ easily inferred.

In order for the bulk space-time to have a holographic dual on the
time-like boundary $B$, however, a Gibbons-Hawking term on this boundary surface
should be added to the action (\ref{eq:action}), as well as a counter term for holographic renormalization should be included.
In this case, we require that the vector field $\xi$ should be tangent
to $B$ and consider the action
\begin{equation}\label{Gibbons-Hawking}
	I=\int_\mathrm{bulk}\mathscr{L}\sqrt{-g}d^{d+1}x+\int_{B}\frac{K}{8\pi G}\sqrt{-\bar{g}}d^d\bar{x}+\int_{B}\mathscr{L}_{CT}\sqrt{-\bar{g}}d^d\bar{x}
\end{equation}
with the Gibbons-Hawking term and the counter term, where $\mathscr{L}$ is the original Lagrangian given in (\ref{eq:action}) and $K$ is the trace of external curvature of the boundary surface $B$. Using variation arguments\cite{TWZ2} with the diffeomorphism invariance of the above action, one can obtain the balance relation
\begin{equation}\label{balance}
	F_f-F_i-W+D=0
\end{equation}
on-shell, where
\[
F=-\int_{\Sigma}m_{\mu}J_\xi^{\mu}+\int_{\Sigma\bigcap B}\bar{m}_\mu (\frac{K}{8\pi G}+\mathscr{L}_{CT})\xi^\mu
\]
the renormalized grand potential (with $\Sigma=\Sigma_f$ and $\Sigma=\Sigma_i$ corresponding to that of the final state and the initial state, respectively) with $\bar{m}_\mu$ the future directed unit normal vector to $\Sigma\bigcap B$ within $B$,
\begin{equation}\label{ren_work}
	W=-\int_{B}(\frac{1}{2}\tilde{t}^{ab}\mathcal{L}_{\xi}\tilde{\bar{g}}_{ab}+\tilde{\pi}_{A}\mathcal{L}_{\xi}\tilde{\bar{\phi}}^{A})\sqrt{-\bar{g}}d^{d}x
\end{equation}
the renormalized external work done to the boundary system with $\pi_{A}=n_{\mu}\frac{\partial\mathscr{L}}{\partial\nabla_{\mu}\phi^{A}}$
the conjugate moment to $\phi^{A}$ and
\begin{equation}\label{dissipation}
	D=-\int_{H}\lambda_{\mu}J_\xi^{\mu}
\end{equation}
has the rough meaning of energy dissipation. Note that in the renormalized external work (\ref{ren_work}), an over bar denotes the boundary
value (or pullback) of the bulk field and the tilde quantities are the renormalized or rescaled ones.\footnote{See also Sec.~IVC in \cite{TWZ2}, but note that in the back-reacted case the term $2{\delta I_{\mathrm{CT}}}/{\delta\bar{g}_{ab}}$ there is itself a renormalization of the boundary stress-energy tensor $t_{ab}$, so it is already included in the term $\tilde{\pi}\mathcal{L}_\xi\bar{\phi}$.}

{Actually, from the momentum constraints
\[
\bar{\nabla}_{a}t^{ab}=-n_{\mu}T^{\mu\nu}
\]
we have
\begin{equation}\label{work_density}
\bar{\nabla}_{a}(t^{ab}\xi_{b})=t^{ab}\bar{\nabla}_{a}\xi_{b}-n_{\mu}T^{\mu b}\xi_{b}=\frac{1}{2}t^{ab}{\cal L}_{\xi}\bar{g}_{ab}+\pi_A {\cal L}_{\xi}\bar{\phi^A}-\bar{\nabla}_{a}(j^{a}{A}_\xi),
\end{equation}
where the second equality holds due to $n_\mu\xi^\mu=0$ and our discussion in Sec.~\ref{probe}. So
\[
W=-\int_B (\bar{\nabla}_{a}[t^{ab}\xi_{b}]+\bar{\nabla}_{a}[j^{a}{A}_\xi])\sqrt{-\bar{g}}d^{d}x=\int_{\Sigma\bigcap B} (\varepsilon-\rho\mu)\sqrt{\gamma}d^{d-1}x\Big|_{i}^{f}
\]
with $\varepsilon=\bar{m}_{a}t^{ab}\xi_{b}$ the energy density, $\rho=-\bar{m}_a j^a$ the particle number density and $\mu={A}_\xi$ the chemical potential (under the gauge $A_\mu|_H=0$), which means that $W$ is equal to the difference of the total ``energy'' of the boundary system between the final state and the initial state, i.e. the external work done to the boundary system under given chemical potential, if $-t^{ab}\xi_b$ is taken as the stress energy current of the boundary system.\footnote{In our discussion, $\xi$ even does not need to be time-like on the boundary, which is the case when the bulk black hole has angular momentum (like Kerr). In that case, $W$ is a combination of work and impulse for the boundary system.}}

{For equilibrium states or the asymptotic Killing regions like $\Sigma_i$ and $\Sigma_f$, it can be generally shown that the free energy $F$ (defined by $K_\xi$ instead of $J_\xi$) and ``internal energy''
\begin{equation}
E=\int_{\Sigma\bigcap B} \varepsilon\sqrt{\gamma}d^{d-1}x
\end{equation}
are related by the Legendre transform (see, e.g. \cite{TWZ2})
\begin{equation}\label{energy_free}
F=E-TS.
\end{equation}
But on an arbitrary time slice under dynamic evolution, the above relation is not expected to hold any longer, since there should not be local correspondence between the boundary and the horizon in holography.}

In a dynamic process of an isolated system, the entropy increases monotonically, but the free energy of the final equilibrium state would not be greater than its initial value (if well defined) in general. So it is expected from the balance relation (\ref{balance}) (with respect to $K_\xi$ instead of $J_\xi$) that $D$ is not always positive if we turn off $W$ (driving). In addition, the vector field $\xi$ remains rather arbitrary. It is not clear whether there is a superior choice of $\xi$, especially near the horizon, which makes the conserved currents most meaningful in the bulk.\footnote{In practice, it is a convenient choice that $\xi$ is just taken as $\partial_t$ with $t$ the time coordinate in dynamic evolution, if it has the asymptotic Killing property.} Nevertheless, our discussion clarifies the external work in fully back-reacted holography as the flux of the stress energy current $K_\xi$ across $B$ and provides possible routes to define physical quantities during the whole dynamic evolution.

\section{Discussion}\label{discuss}

{In this paper, we have studied dynamical holographic systems and the relation between thermodynamical and dynamical stability of such systems, with the help of the conserved currents associated to the diffeomophism invariance. It is noteworthy that the similar idea has been used to prove the equivalence of dynamical and thermodynamical stability for pure gravity in \cite{Wald2}, where the Noether current is used instead of our stress energy current. So it is definitely interesting to see the relation between our discussion and that in \cite{Wald2}. We hope to address this issue in the future.}

{Our analysis in the probe limit is very clear, where a generalized free energy with very nice properties in fully dynamical holographic systems is defined. However, in the back-reacted case, although we have some discussion based on the conserved currents corresponding to the diffeomorphism as well, there are a lot of open problems to be solved. Due to the conceptual difficulties in the fully back-reacted case, as the next step we may consider the gravitational dynamics at the linear perturbation level, aiming at the linear stability analysis similar to the probe-limit case. At this level we can use the effective action of the gravitational perturbation $h_{ab}$ at quadratic order, and then the method in Sec.~\ref{probe} should give similar results of the conserved currents and the relations between them due to the linearized diffeomophism transformation $h_{ab}\to h_{ab}+\nabla_a\zeta_b+\nabla_b\zeta_a$ as a gauge symmetry.}

{In Sec.~\ref{canonical_dynamics}, we have demonstrated how to set up a holographic system (in the probe limit) under the environment of a particle source with fixed chemical potential. It is expected that in the back-reacted case we can deal with an isolated holographic system (with fixed total energy) or a holographic system under the environment of a heat sink with fixed temperature similarly, by imposing different spatial boundary conditions for the metric fields respectively, though a thorough discussion on this aspect still needs a lot of work.}

\begin{acknowledgments}
We would like to thank {Bob} Wald for referring to the reference \cite{Wald2}. Y.T. is grateful to Xin Li and Hong Liu for helpful discussions. This work is partially supported by NSFC with Grant No.11975235, No.12035016, No.12075026 and No.12275350. H.Z. is supported in part by the Belgian Federal
Science Policy Office through the Interuniversity Attraction Pole
P7/37, by FWO-Vlaanderen through the project G020714N, and by the Vrije Universiteit Brussel through the
Strategic Research Program ``High-Energy Physics''. He is also an individual FWO fellow supported by 12G3515N.
\end{acknowledgments}


\begin{thebibliography}{1}
\bibitem{Sonner}H. Liu and J. Sonner, [arXiv:1810.02367].
\bibitem{AdS-CFT}J. M. Maldacena, Adv. Theor. Math. Phys. 2, 231(1998).
\bibitem{Witten}E. Witten, Adv. Theor. Math. Phys. 2, 253(1998).
\bibitem{GKP}S. S. Gubser, I. R. Klebanov and A. M. Polyakov, Phys. Lett. B 428, 105(1998).
\bibitem{Minkowski}D. T. Son and A. O. Starinets, JHEP 0209, 042(2002) [hep-th/0205051].
\bibitem{GM1}S. S. Gubser and I. Mitra, [hep-th/0009126].
\bibitem{GM2}S. S. Gubser and I. Mitra, JHEP 08, 018(2001) [hep-th/0011127].
\bibitem{Wald1}V. Iyer and R. Wald, Phys. Rev. D 50, 846(1994).
\bibitem{Reichl}R. E. Reichl, \textit{A Modern Course in Statistical Physics}, University of Texas Press, Austin, TX, 1980.
\bibitem{vortex}P. M. Chesler, H. Liu and A. Adams, Science 341(6144): 368.
\bibitem{TWZ1}Y. Tian, X. Wu, and H. Zhang, Class. Quant. Grav. 30, 125010(2013).
\bibitem{TWZ2}Y. Tian, X. Wu, and H. Zhang, JHEP 10, 170(2014).
\bibitem{driving}W.-J. Li, Y. Tian and H. Zhang, JHEP 07, 030(2013).
\bibitem{HHH1}S. A. Hartnoll, C. P. Herzog and G. T. Horowitz, Phys. Rev. Lett. 101, 031601(2008).
\bibitem{HHH2}S. A. Hartnoll, C. P. Herzog and G. T. Horowitz, JHEP 12, 015(2008).
\bibitem{LanTZ}S. Lan, Y. Tian and H. Zhang, JHEP 07, 092(2016) [arXiv:1605.01193].
\bibitem{Wald2}S. Hollands and R. Wald, Commun. Math. Phys. 321, 629(2013) [arXiv:1201.0463].
\bibitem{DNTZ}Y. Du, C. Niu, Y. Tian and H. Zhang,  JHEP 12, 018(2015) [arXiv:1412.8417].
\bibitem{DLTZ}Y. Du, S.-Q. Lan, Y. Tian and H. Zhang, JHEP 01, 016(2016) [arXiv:1511.07179].
\bibitem{LTZ}X. Li, Y. Tian and H. Zhang, JHEP 02, 104(2020) [arXiv:1904.05497].
\bibitem{Chesler}P. M. Chesler and L. G. Yaffe, JHEP 07, 086(2014) [arXiv:1309.1439].
\bibitem{NCTWZ}Z. Ning, Q. Chen, Y. Tian, X. Wu and H. Zhang, [arXiv:2307.14156].
\end{thebibliography}
\end{document}